\documentclass[aps,preprintnumbers,twocolumn,amsmath,amssymb,prl,showpacs]{revtex4}
\usepackage{bbm}
\usepackage{mathrsfs}
\usepackage{amsmath}
\usepackage{graphicx}
\usepackage{epstopdf}
\usepackage{ulem}
\usepackage{color}

\newcommand{\be}{\begin{equation}}
\newcommand{\ee}{\end{equation}}
\newcommand{\bea}{\begin{eqnarray}}
\newcommand{\eea}{\end{eqnarray}}
\newcommand{\bes}{\begin{split}}
\newcommand{\ees}{\end{split}}

\newcommand{\tr}{\operatorname{Tr}}

\begin{document}
\title{Deep learning of topological phase transitions from entanglement aspects}
\author{Yuan-Hong Tsai$^{1}$}\email{yhong.tsai@gmail.com},
\author{Meng-Zhe Yu$^{2}$}
\author{Yu-Hao Hsu$^{2}$} 
\author{Ming-Chiang Chung$^{2,3,4}$}\email{mingchiangha@nchu.edu.tw} 
\affiliation{$^1$ Taiwan AI Academy, New Taipei, 241, Taiwan}
\affiliation{$^2$ Physics Department, National Chung-Hsing University, Taichung, 40227, Taiwan}
\affiliation{$^3$ Max-Planck-Institute for Physics of Complex Systems, N{\"o}thnitzer Straße 38 01187, Dresden, Germany  }
\affiliation{$^4$ Physics Department, Northeastern university, 
360 Huntington Ave., Boston, Massachusetts 02115, U.S.A. }

\begin{abstract}
The one-dimensional $p$-wave superconductor proposed by Kitaev has long been a classic example for understanding topological phase transitions through various methods, such as examining Berry phase, edge states of open chains and, in particular, aspects from quantum entanglement of ground states. In order to understand the amount of information carried in the entanglement-related quantities, here we study topological phase transitions of the model with emphasis of using the deep learning approach. We feed different quantities, including Majorana correlation matrices (MCMs), entanglement spectra (ES) or entanglement eigenvectors (EE) originated from Block correlation matrices (BCMs), into the deep neural networks for training, and investigate which one could be the most useful input format in this approach. We find that  ES is  indeed too compressed information compared to MCM or EE. MCM and EE can provide us abundant information to recognize not only the topological phase transitions in the model but also phases of matter with different $U$(1) gauges, which is not reachable by using ES only.  
\end{abstract} 
\pacs{}

\date{\today}
\maketitle 

{\it Introduction--}
Going beyond Ginzburg-Landau theory of phase transitions \cite{GinzburgLandau}, a topological phase transition (TPT) can occur when no symmetry is broken in a physical system. Since the discovery of integer quantum Hall effect \cite{IntegerQH}, the very first example of TPTs, various materials potentially exhibiting TPTs have been proposed in recent years, although only some of them have been confirmed experimentally \cite{Molenkamp07, Bansil16}. Among them, the one-dimensional topological $p$-wave superconductor proposed by Kitaev \cite{kitaev01} has become one of the most interesting proposals due to the fact that the edge modes in these superconductors can be viewed as ``Majorana fermions'', whose anti-particle is the particle itself. In particular, they are essential components in forming practical fault-tolerant quantum computers \cite{nayak08}. To realize such superconducting state, clever combinations of topological materials or semiconductors with an ordinary  $s$-wave superconductor through proximity effect have been proposed \cite{FuKane07}. In the past few years, Majorana fermions/zero modes have been claimed found at the edges of certain systems either by directly observing the STM spectroscopy \cite{Mourik12, Yazdani14, Albrecht16} or indirectly by measuring the 4$\pi$-periodicity of Josephson junction currents \cite{Wiedenmann16}.

In fact, Majorana zero modes can be alternatively detected by measuring quantum entanglement, a concept in the subject of quantum information. For instance, the von Neumann entanglement entropy of a subsystem $A$ embedded in the environment $B$ can be obtained via the density matrix after tracing out the degrees of freedom in $B$, {\it i.e.}, the sub-entropy $S_A = - \tr \rho_A\log_2 \rho_A$ with the reduced density matrix $\rho_A = \tr_B |\Psi_{A \cup B} \rangle \langle\Psi_{A \cup B} |$ \cite{BCMReview,BCMa,BCMb,BCMc,BCMd}. To reveal the topological nature of the system, one can further analyze the eigenvalues of the $\rho_A$-deduced entanglement Hamiltonian: The presence of Majorana zero (edge) modes can now be inferred by a corresponding degeneracy in the entanglement spectrum. In other words, the spectrum preserves topological information. Moreover, for any system with a quadratic Hamiltonian, the whole situation can be simplified to compute the eigenvalues of the correlation function matrix, also known as the one-particle entanglement spectrum \cite{hatsugai06,chung16,Fromholz}. Majorana zero modes are then shown in terms of doubly degenerate eigenvalues, $1/2$, in the spectrum \cite{chung16}. This example  reflects the power of the entanglement-related quantities as a good topological indicator, but sometimes computation of these quantities is time-consuming and measuring entanglement is experimentally difficult. And hence, a more efficient tool for identifying TPTs might be necessary.

Machine learning (ML) is a rapidly growing field of computer science recently due to the availability of large-scaled datasets and advances in the computation hardware. Its applications have become ubiquitous in our daily life from automated machine translation, vision and speech recognition, matching news items, to email spam-filter and so on \cite{lecun15}. By feeding in a large amount of data (or ``features''), ML algorithms can ``learn'' to condense them into a more accessible/meaningful form, such as distinguishable classes or patterns. In particular, a neural network-based learning method, called deep learning (DL), are composed of several simple but non-linear modules and able to effectively learn suitable representations from complex raw data and distill essential information. Therefore, as a straightforward application on quantum matters, so far DL has been employed by physicists to classify different phases of matter and identify phase boundaries \cite{Nieuwenburg17,Carrasquilla17,Ohtsuki16,wang16,Broecker17, Scheurer}. Moreover, the remarkable property of this method is also shown in considering topological \cite{kim17a, kim17b} or out-of-equilibrium systems \cite{kim18} where no obvious local order parameter is available.

Despite recent progress in using DL for identifying phase transitions, many of them are based on straightforward wave functions (by certain manipulation or not) as the input data for learning, while relatively few are based on the aspect of quantum information. Since quantum information is also known to be useful when there is no local order parameter available in a system, one pioneering work takes the entanglement spectrum (ES), used to compress the ground state information, as the input data and trains a neural network to distinguish the topological phase from the trivial one \cite{Nieuwenburg17}. However, one should ask if ES is too compressed for a given quantum system and are there better ways to represent quantum information for the purpose of using DL technique. Therefore, in this work, we study topological phase transitions in 1D $p$-SC via DL approach and systematically examine what kinds of quantum information related quantities could better represent the features of a quantum system, and hence a better input for DL. Concretely, we find that block correlation matrix in Majorana representation not only tip out the phase boundaries but can even offer extra information beyond ES.

{\it Model --}  The $p$-wave superconducting system of spinless fermions \cite{kitaev01} in one dimension (1D) is described by the Hamiltonian  
\bea  
H &= & \sum_{i}  -t \left (c_i^{\dagger} c_{i+1} + c_{i+1}^{\dagger} c_i \right)  \nonumber \\ 
   &+& \Delta \left(c_i c_{i+1} + c_{i+1}^{\dagger} c_{i}^{\dagger} \right)  - \mu
  \left( c_{i}^{\dagger} c_i -1/2\right),  
\label{H:pwave}
\eea
where $t$ is the nearest-neighbor hopping amplitude, $\Delta$ is superconducting pairing potential, and $\mu$ represents on-site chemical potential. With the translational invariance, Eq.~(\ref{H:pwave}) can be rewritten as
\be  \label{H:R} 
H = -\sum_{k \in BZ} \left (c_k^{\dagger}, c_{-k} \right ) 
\left[ {\mathbf R}(k) \cdot{\boldsymbol \sigma} \right] 
\left (c_k, c_{-k}^{\dagger} \right )^T, 
\ee 
where Pauli matrices ${\boldsymbol \sigma} = (\sigma_x, \sigma_y, \sigma_z)$, and ${\mathbf R}(k) =  (0, -\Delta \sin{k}, t\cos{k} + \mu/2)$ is the pseudo-magnetic field. The one-particle energy spectrum is simply 
$\epsilon (k) = \pm 2 R(k) = \pm \sqrt{(2t \cos{k} + \mu)^2 + 4\Delta^2 \sin^2{k}} $. 
Note that this system preserves particle-hole symmetry, while it breaks time reversal symmetry and hence, the chiral symmetry; therefore it belongs to the class D according to the ten-fold way classification for symmetry-protected topological systems \cite{Schnyder08}. It can be characterized by a $Z_2$ topological invariant. 

\begin{figure}[t]
	\begin{center}
		\includegraphics[width=7.5cm]{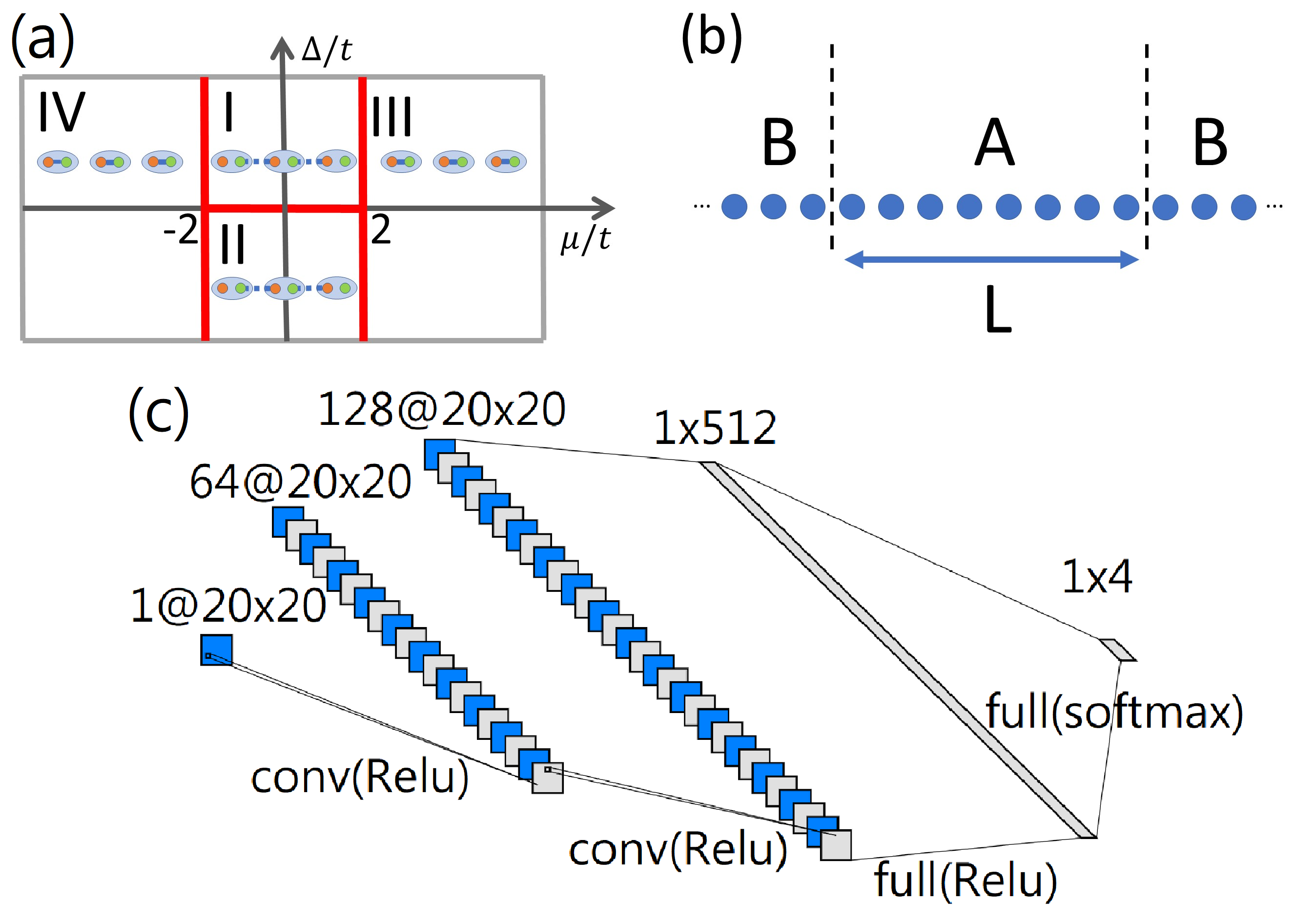}
			\caption{(Color online) (a) Topological phase diagram of the 1D $p$-wave superconductor, where the chain-like inset pictures show schematically the Majorana representation for each phase as described in the main context. These indicate phases I and II are topological, while the others are not. (b) The infinite system is divided into a finite subsystem A with $L$ sites and an environment B. (c) The schematic illustration of the convolutional neural network used in this work.}
		\label{fig:network}
	\end{center}
\end{figure}

To make physics more transparent, we define Majorana operators, $d_{2j-1} = c_j + c_j^{\dagger}$ 
and $d_{2j} = -i(c_j  - c_j^{\dagger})$, and Eq.~(\ref{H:pwave}) becomes 
\bea
H &=& \frac{i}{2} \sum_j [(-t+|\Delta|) d_{2j-1} d_{2j+2} + (t+|\Delta|) d_{2j}d_{2j+1}  \nonumber  \\
&-& \mu d_{2j-1} d_{2j}].
\eea
As shown in Fig. 1(a), when $|\mu| > 2t$, the system Hamiltonian can be adiabatically connected to the form, $\frac{-i\mu}{2}\sum_j d_{2j-1}d_{2j}$, where $t=|\Delta| =0$ and $\mu < 0$. It is then straightforward to see that the ground state is now composed of a paired Majorana fermions at the same site, resulting in no Majorana edge modes and hence a topologically trivial phase (phases III and IV).  On the other hand, when $|\mu| < 2t$, $H$ is adiabatically connected to the special case, $it \sum_j d_{2j} d_{2j+1}$, where  $t=|\Delta| >0$ and $\mu = 0$: Most Majorana fermions from neighboring sites are paired together while the system leaves the edge Majorana modes alone (unpaired), and thus corresponding to a nontrivial phase (phases I and II).

As we have mentioned in the introduction, quantum information of the system can serve as a useful diagnostic for topological phase transitions, and in particular, the entanglement correlations should encode whole information about the focused state. In this paper, we consider two common correlators. First, by separating the system into A and B blocks [see Fig. 1(b)] and in terms of Majorana operators, the Majorana correlation matrix (MCM) of Majorana fermions at different sites within the subsystem A can be defined as $\text{MCM}_{i,j} \equiv i \tr \rho_0 d_{2i} d_{2j-1} = \tr \rho_0 (c_i - c_i^{\dagger})(c_j +c_j^{\dagger})$, 
where $\rho_0$ represents the density matrix of the ground state. The other correlations are either the conjugation or proportional to trivial identity when system Hamiltonian is quadratic. Second, the block correlation matrix (BCM) for subsystem A is defined as 
$\text{BCM}_{i,j} = \tr \rho_0 {\hat{\mathbf{c}}}_i \hat{{\mathbf{c}}}_j^{\dagger} $ 
with $\hat{{\mathbf{c}}}_i \equiv (c_i, c_i^{\dagger})^T $ and $i,j$ being sites of the finite block A. This matrix is intimately connected to the more familiar quantity, the reduced density matrix of the block A,
$\rho_A = \bigotimes_m
\left[\begin{matrix} \lambda_m & 0\\ 0 & 1-\lambda_m \end{matrix}
\right]$, 
where $\lambda_m$ are simply the eigenvalues of BCM and $\lambda_m$s are also known as one-particle entanglement spectrum (OPES). Therefore, the eigenvalues and their corresponding eigenvectors of BCM could be considered as potential representations of quantum information in the system for our deep learning purpose. 

{\it DL-based approach --}  In order to correlate the phases outlined in the previous section with various quantum information-inspired input representations, we employ convolutional neural networks (CNNs), which are non-linear functions particularly designed for efficiently recognizing patterns in the image-type data \cite{LeCun89,LeCun98}. As we explain below in more detail, our choice of using CNNs is based on a natural interpretation of the input representations as ``images''. 

Using Keras \cite{keras}, we build a deep CNN architecture as depicted schematically in Fig. 1(c). The architecture is mainly composed of two parts: Begins with convolutional layers followed by fully-connected neural networks. The convolution part processes the input data by two consecutive convolutional layers both with filters of kernel size $3\times 3$ and rectified linear unit (Relu) activation functions. The number of filters (depth) is 64 for the first layer, and 128 for the second one. We do not insert pooling layers here to avoid missing subtle information due to our small image size (usually smaller than $100\times 100$). After convolution, the processed data is then fed into a classifier made of a fully-connected, Relu activated layer with 512 neurons and a four-neuron, fully-connected softmax layer. The final outputs after softmax activation would sum up to unity and thus can be interpreted as the probabilities that the input data belongs to the four different phases as shown in Fig.~1(a).

\begin{figure}[th]
	\begin{center}
		\includegraphics[width=7.5cm]{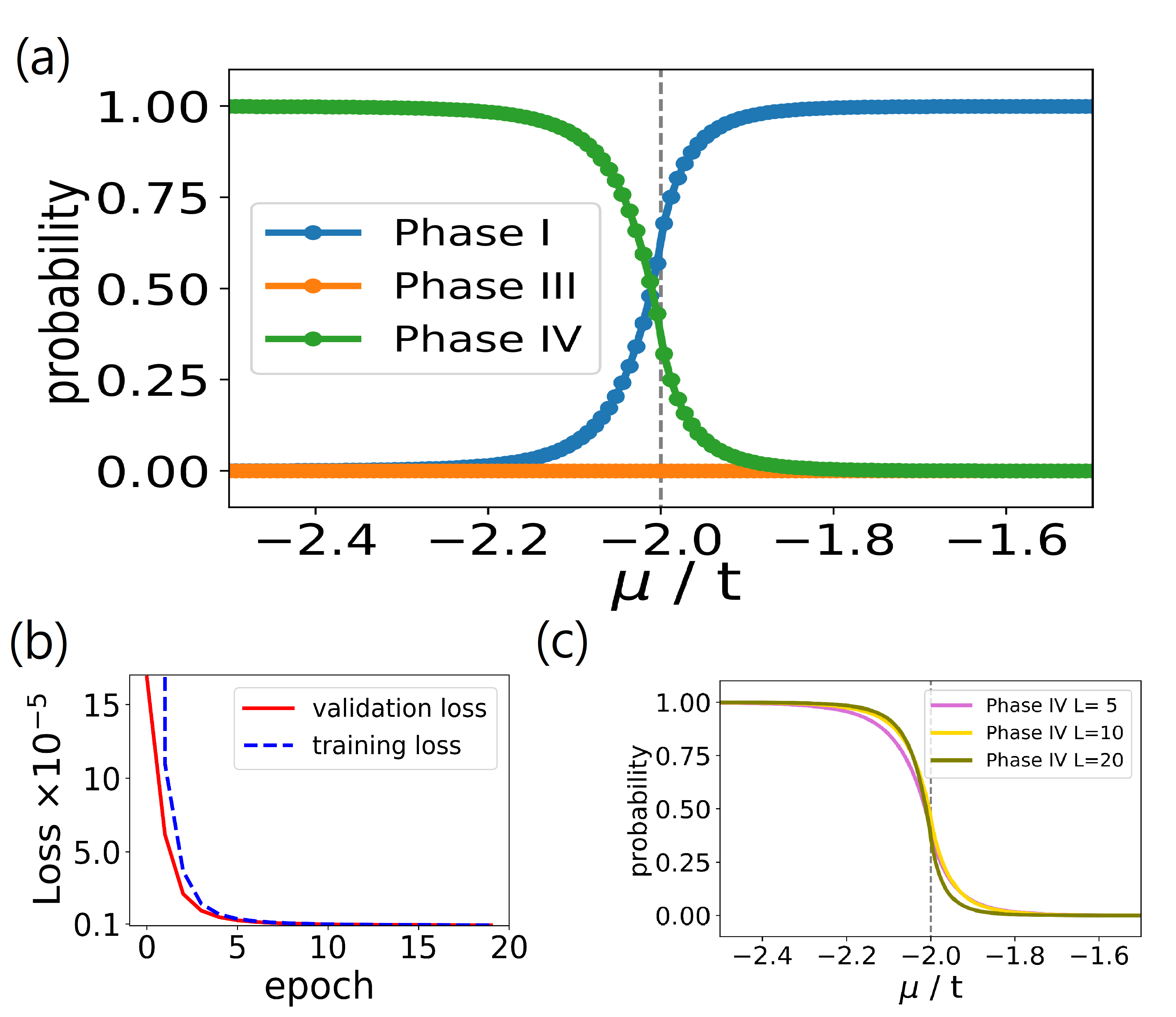}
		\caption{(Color online) (a) Each neuron output of the final softmax layer, corresponding to the probability of each phase, as a function of $\mu/t$ (unseen data) with $\Delta/t=1, L=20$. Although the training sets from MCMs are far beyond the $\mu/t$ region shown here, the CNN can still recognize a topological phase transition near -2.0. The dashed line indicates the theoretical value of the transition. (b) The validation loss follows the trend of the training loss well, suggesting no over-fitting happened. (c) The transition step becomes sharper as $L$ grows,  showing the finite-size effect of the subsystem A.}
		\label{fig:MCM}
	\end{center}
\end{figure}

At the supervised training stage, we train the CNN on a dataset composed of $\mathcal{O}(10^4)$ representative ``images'' in ``gray scale'' (typically of $20\times 20$ or $40\times 40$ pixels) generated simply around six $\mu$ values (in a width of $0.05t$), corresponding to four possible phases deep inside the phase diagram [See Figs.~1(a) and 3]. Note that we collect ``image'' data in each {\it topological} phase around two different $\mu$ values for training to inform the model about two possible phase boundaries. Setting the train-validation split ratio as 0.2, the optimization for our model is then performed by ADAM algorithm \cite{adam} at learning rate $10^{-3}$ with cross entropy as the loss function. Typically after training over 15 epochs, both train and validation losses would be less than $10^{-6}$, indicating that the resultant model becomes reliable [See Fig.~2(b)]. Once training is done, at the inference stage we fix whole parameters in the trained model and feed with new data for prediction.

{\it Results --}  Now, we take aforementioned deep learning approach to study topological phase transitions occurred in 1D $p$-SC and examine various quantum information-inspired input features in order to provide a better compressed representation of the naive ground state wave function. We first prepare the training input ``images'' with labels by calculating a thousand of MCMs with subsystem size $L$ (block A) under periodic boundary conditions of the full system around a given $\mu$ for each phase. Each MCM can be viewed as a $L\times L$ ``image'' in one (gray) channel and entries in a MCM represent pixel values. For simplicity, we fix $\Delta=t$ in the training set (except for phase II, where $\Delta=-t$). After training, we find that the model easily learns how to distinguish different phases for the given dataset and is ready to generalize for unseen data points.

\begin{figure}[thb]
	\begin{center}
		\includegraphics[width=7.5cm]{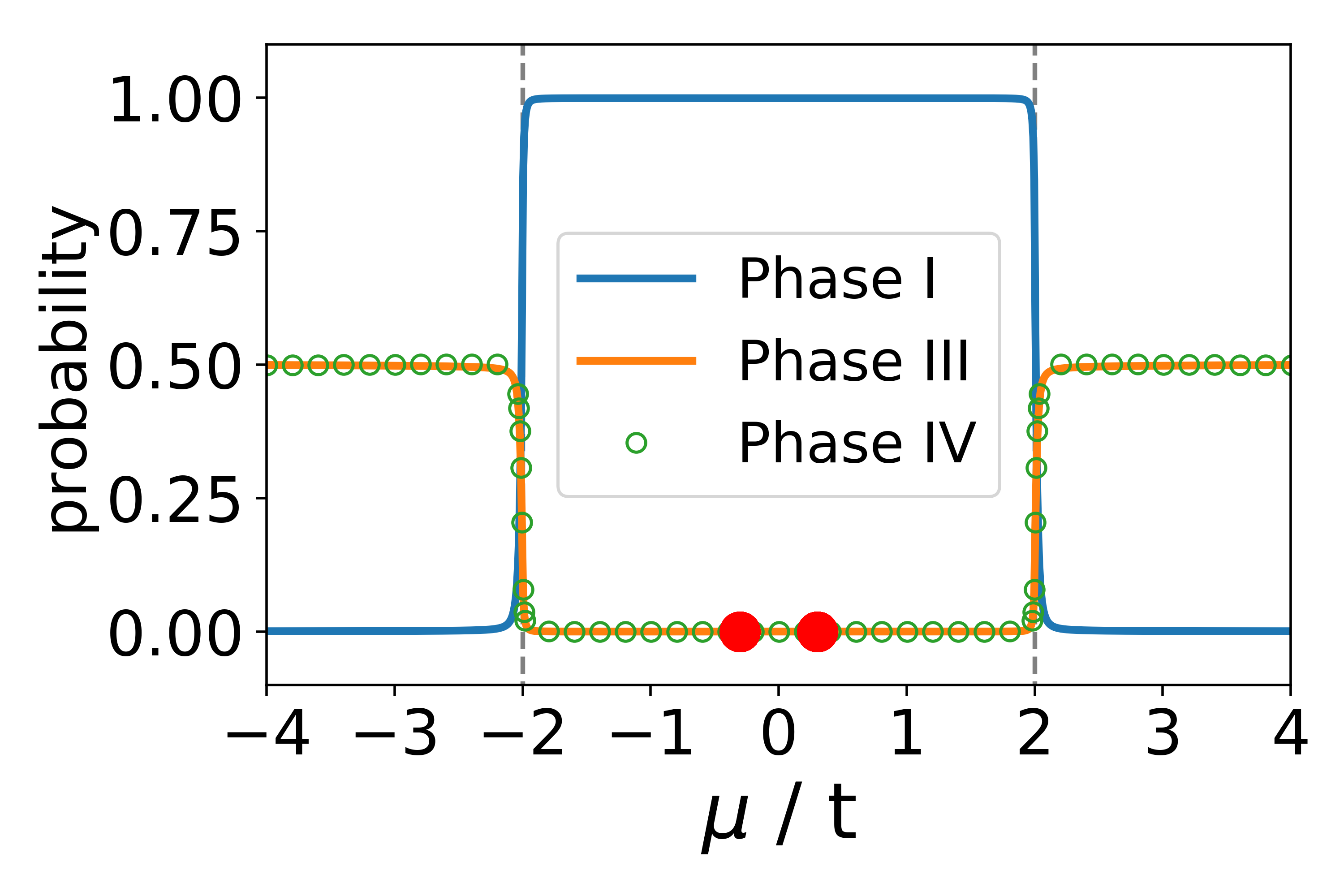}
		\caption{(Color online) The neuron output ``phase diagram'' is shown as a function of $\mu/t$ with $\Delta/t=1, L=20$. The training sets from the eigen-spectra of BCMs are prepared at around $\mu/t=-13, -0.3, 0.3, 13$ of a window width 0.05 (red dots). The phase boundaries are clearly recognized for unseen data points, but the probabilities corresponding to phases III and IV are equal at $|\mu/t|>2$, suggesting they confuse the neural network.}
		\label{fig:BCM_value}
	\end{center}
\end{figure}

As shown in Fig. 2(a), the neuron output corresponding to phase IV goes from probability 1 at $\mu/t=-2.5$ to 0 at $\mu/t=-1.5$. Such curve gets across at $\mu^{*}(L=20)/t\approx -2.01$ with the curve corresponding to phase I which behaves just oppositely, indicating that our CNN model indeed realizes a phase transition. In addition, the curves corresponding to the other phases are never activated here. Similarly, at $\mu^{*}(L=20)/t\approx 2.01$ two curves corresponding to phases I and III, respectively, also cross with each other and thus again it suggests another occurrence of the phase transition (not shown). 

Note that for a given finite-size $L$, $\mu^{*}(L)$ is identified as the point where both crossing curves have equal probability 0.5, {\it i.e.}, at the moment that our trained model is not able to distinguish between the two phases. The non-abruptness of the phase transition seen from the probability curves is due to finite-size effect. As one can see in Fig. 2(c), the transition region becomes sharper when $L$ grows longer and the finite-size trend of $\mu^{*}(L)/t$ is also closer to $\mu^{*}/t=-2$ in the thermodynamic limit where $L\rightarrow\infty$.

\begin{figure}[th]
	\begin{center}
		\includegraphics[width=7.5cm]{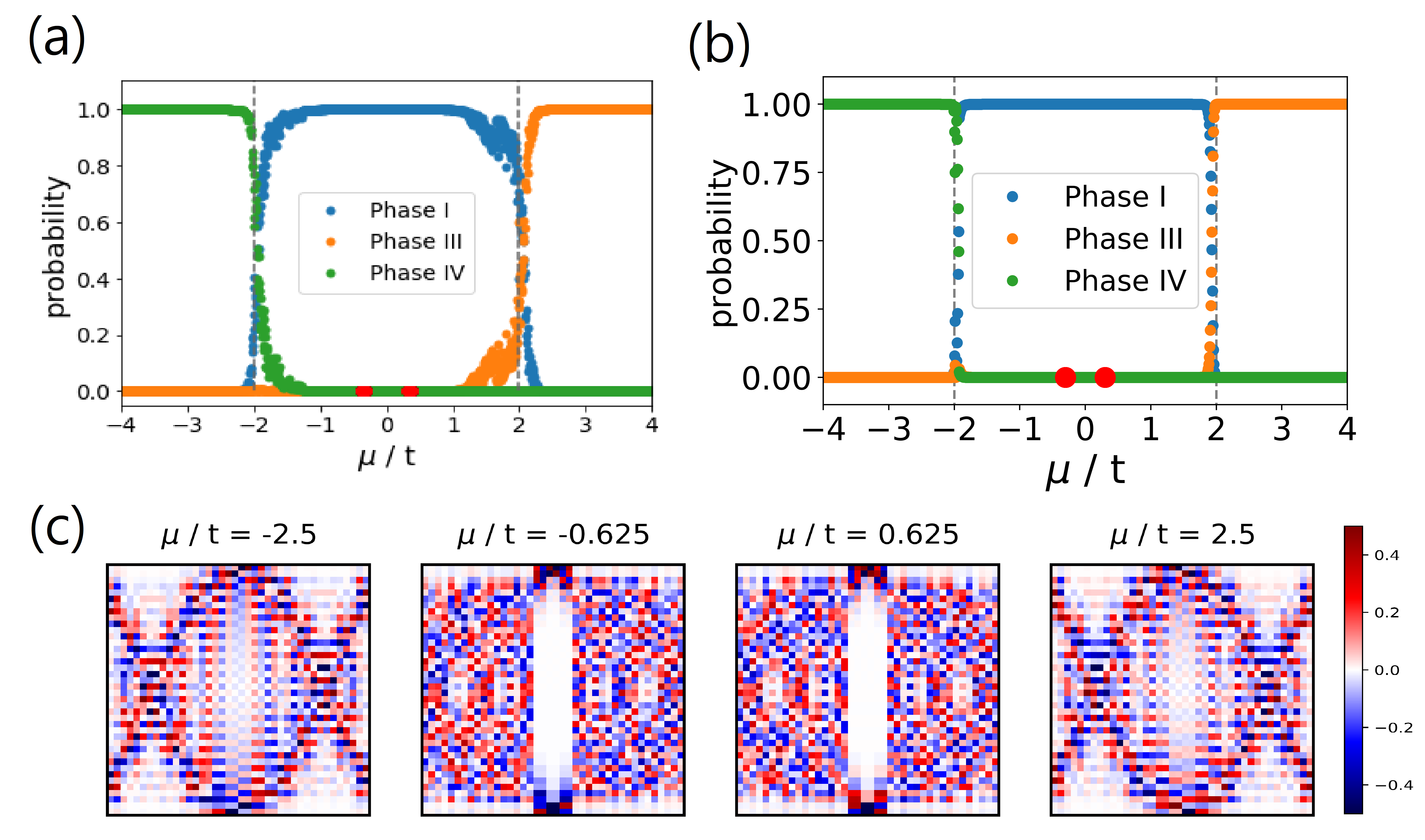}
		\caption{(Color online) (a) The neuron output ``phase diagram'' is shown as a function of $\mu/t$ with $\Delta/t=1, L=20$. The training sets from whole eigenvectors of BCMs are prepared at the same $\mu/t$ regions mentioned in Fig. 3 (red dots). The phase boundaries are still sharply defined as the MCM calculations. (b) Similar diagram as (a) but with only two middlemost eigenvectors making the input ``images''. The phase boundaries are sharply shown. (c) Representative input ``images'' for the CNN at given $\mu/t$ values in the case (a). Boundary modes or topology-induced end modes are present at the middle-top and the middle-bottom parts of each image.}
		\label{fig:BCM_vector}
	\end{center}
\end{figure}

Alternatively, we next consider taking BCM as our training input. For a finite subsystem A of size $L$, there are at least three ways to represent each BCM (now of size $2L\times 2L$ due to Nambu notation) as an ``image'': (i) View BCM itself as a ``gray image''; (ii) arrange all eigenvalues of the BCM in ascending order into a diagonal matrix, which can be viewed as an ``image''; (iii) we diagonalize a BCM and then arrange each eigenvector as one of the columns in a new matrix $M$. $M$ is again of size  $2L\times 2L$ and can be viewed as a ``gray image''. In the DL approach, each way of representing the focused ground state would encode different levels of entanglement information, likely leading to distinguishable ability in realizing different phases, which is what we want to examine now. Since by definition of MCM, way (i) of treating BCM should be similar to the MCM case and hence we do not repeat it here and only focus on the latter two ways.

In fact, in way (ii) utilizing diagonal matrices of BCMs from all eigenvalues as input ``2D images'' is not efficient because the off-diagonal part contains no information (like black background). Therefore, we simply feed in all eigenvalues of a BCM, namely, entanglement spectrum, as an 1D input to a simple feed-forward neural network composed of 3 consecutive fully-connected layers (with Relu activation), having 32, 64, 256 neurons, respectively, followed by a 4-neuron dense layer with softmax activation as the final output. By taking the same training procedure (skip phase II training samples) and waiting until the training and validation losses converge, we show the final predictions of unseen data points at $\Delta/t =1$, $L=20$ as a function of $\mu$ in Fig. 3. It is clear that phase I can be recognized very well, while the outputs corresponding to phases III and IV are not. The predicted probabilities are 0.5 and 0.5, respectively, indicating that these two phases confuse the network due to their similar eigen-spectra. If one examines phases III and IV carefully, one can see that they indeed belong to the same superconducting phase but very likely with different $U$(1) gauges. The overall results for this case thus suggest that the representation using entanglement spectrum as inputs compresses too much quantum information and may not be effective when considering to determine the global phase diagram via DL approach.

To gain more quantum information, we finally investigate way (iii) to prepare the input dataset for training the slightly different CNN model. We keep the first convolution and then connect with two consecutive residual blocks. Then the global average 2D is used before final dense layer-based classifier {\cite{He}}. Once complete training, Figs. 4(a) and 4(b) depict again the model predictions for the unseen data points as a function of $\mu$ at $\Delta/t=1$, $L=20$. In the case of Fig. 4(a), we make each input ``image'' by including all eigenvectors of each BCM, while by including only two middlemost eigenvectors in the case of Fig. 4(b). Clearly, both cases show the ability of the CNN model to recognize global phase diagrams despite of different $U$(1) phases which phases III and IV may take, and similar results are also observed when the unseen data points vary along $\Delta$ with fixed $|\mu|\le 2$ (i.e. phases I and II). But one would notice that the Fig. 4(a) case is slightly worse than 4(b) due to more eigenvectors, which may be considered as noises.  

The underlying reason behind the above result can be explained as follows. Let us examine some selected input ``images'' made of all eigenvectors at $\mu/t=2.5, -0.625, 0.625, 2.5$, respectively, in Fig. 4(c). The essential feature to distinguish a nontrivial topological phase from the trivial one is the presence of end modes (with entanglement eigenvalue 0.5) in the middlemost region of the figure along $x$. Thus, importantly, this indicates that our model can differentiate the end modes ($\mu/t=-0.625, 0.625$ cases) from the boundary eigenstates due to the finiteness of the subsystem A ($\mu/t=-2.5, 2.5$ cases), which is difficult by human eyes. However, when taking all eigenvectors in making ``images'', the bulk eigenvectors occupy large portion of the image which may force our model to pay more attention to this portion and hence weaken the ability of the model to determine the phase boundaries.

{\it Discussion and conclusion --} There is compelling reason for an independent check of our
aforementioned results by comparing with those from seemingly unrelated 1D transverse-field Ising model. In fact, the transverse-field Ising model with a magnetic field $\lambda$, written as $H_{tI} = -\sum_{i} \sigma^{x}_i \sigma^x_{i+1} - \lambda \sum_i \sigma^z_i$ with Pauli matrices $\sigma^{\alpha}$, can be transformed into 1D $p$-wave superconductor with $t=-\Delta=1$ and $\mu = 2\lambda$ [see Eq.~(\ref{H:pwave})], by a non-local Jordan-Wigner transformation.  Given the correlation function matrices, the entanglement spectra or entanglement eigenstates as a possible form of inputs, the transition point $\lambda=1$ is still stood out via the proposed deep learning approach to distinguish between the ferromagnetic phase ($\lambda >1$) and the paramagnetic one ($\lambda <1$). It justifies the effectiveness of using the quantum information (in particular, the entanglement aspect) to encode quantum phases in the deep learning process. 

Moreover, among the quantum information-related quantities used in this study, since a $2N \times 2N $ BCM has $N$ redundant variables, we notice that using $N\times N$ MCM would be enough for recognizing topological phase transitions.  And the fact that a relatively small size of MCM is already effective in our approach reflects the nature of BCMs, which indeed includes essential quantum information of the infinite chain at the ground state . 
  
In summary, we demonstrate how to adopt deep learning approach assisted via entanglement aspect to discover topological phase transitions. Several quantum information-related quantities such as MCM, ES or EE (from BCM) are fed into deep neural networks for training. While ES can only find the phase transition points, MCM and EE contain much abundant information to find not only the critical points but also phases of matter with different $U$(1) gauges. Our work emphasizes utilizing quantum information, instead of naive wave functions, as inputs in the deep learning approach and it might be proved useful as well in the higher dimensional systems. 

M.C.Chung acknowledges the MoST support under the contract Nos 108-2112-M-005 -010 -MY3.

\end{document}